\documentstyle[12pt]{article}
\begin{document}
\begin{center}
{\bf \large Full determination of transition matrix elements in the $\vec{d}
(\vec{p},pp)n$ reaction}
\end{center}
\vspace{1cm}
\begin{center}
{F.D.Santos$^*$ and A.Arriaga$^+$}
\end{center}
\begin{center}
\small\it{$^{*+}$Centro de F\'{\i}sica Nuclear da Universidade de Lisboa,
 1699 Lisbon, Portugal}\\
\small\it{$^{*}$Department of Physics, University of Wisconsin, Madison,
WI 53706, USA}
\end{center}
\vspace{3cm}
\begin{center}
Abstract
\end{center}
It is shown that in the $d(p,pp)n$ reaction with a colinear geometry and the
two outgoing protons in a singlet state, the only two independent T-matrix
elements can be determined from spin correlation experiments. These matrix
elements are strongly dependent on the deuteron helicity amplitudes, which
are expected to vanish for $1.4fm^{-1}<p<1.7fm^{-1}$.\vspace{10 mm}\\

\newpage
The deuteron structure is fairly well known for relative distances between
 the neutron and proton larger than $1 fm$, which corresponds to internal
 momenta smaller than $200 MeV/c$. At higher momenta the description of
 the deuteron in terms of two nucleons becomes sensitive to different NN
 interactions and furthermore the effects of other degrees of freedom,
 namely baryon resonances and quarks, are likely to arise. In the momentum
 range of $0.3 - 1.2 GeV/c$ the deuteron helicity amplitudes are expect to
 vanish. The precise location of these nodes should contain information on
 the short-range NN repulsion, and on the non-nucleonic degrees of
 freedom.\\

Measurements of polarization observables can provide new insight into the
 deuteron momentum distribution. In particular, we note that the future
 availability of polarized gas targets makes it possible to perform spin
 correlation experiments. Recent data \cite{ab83}-\cite{ch92} on deuteron
 breakup at intermediate and high energy, reveals a strong sensitivity to
 the high momentum behaviour of the deuteron and the NN interaction. In
 the present letter we consider a particular type of spin correlation
 experiment in the $d(p,pp)n$ reaction, which is of special interest.\\

For a coplanar kinematical configuration, the usual experimental
 configuration, and assuming parity conservation, the number of
 independent transition matrix elements corresponding to the different
 spin projections is 24. This number is reduced to 6 if we detect the two
 outgoing protons with equal momentum, therefore requiring that they are
 in a spin-singlet state. A further reduction to two independent matrix
 elements can be achieved in a colinear experiment where the singlet
 di-proton is detected either in the forward direction $\theta=0^\circ$
 or at $\theta=180^\circ$. In fact with this geometry the sum
 of the spin projections along the incident momentum must be equal in the
 entrance and exit channels. Thus we are left with two independent
 T-matrix elements $< \vec{k}'; \sigma_n |\;T\;| \vec{k};\sigma_d,
\sigma_p>$
\begin{equation}
T_+ = < \vec{k}'; \frac{1}{2} |\;T\;| \vec{k}; 0, \frac{1}{2} >
\label{eq:T+}
\end{equation}
\begin{equation}
T_- = < \vec{k}'; -\frac{1}{2} |\;T\;| \vec{k}; -1, \frac{1}{2} >
\label{eq:T-}
\end{equation}
where $\vec{k}$ is the incident proton momentum and $\vec{k}'$ is the
 di-proton momentum. Here the T-matrix is the solution of the
 Lippmann-Schwinger equation in the 3-particle C.M. frame. Matrix elements
 where all the spin projections $\sigma_p$, $\sigma_d$, $\sigma_n$ reverse
 sign are related to $T_\pm$ by parity conservation.\\

The number of independent spin correlation observables in the
 $\vec{d}(\vec{p},(pp)_\circ)n$ reaction can be easily determined using the
 spin transfer coefficients

\begin{equation}
T_{k_p q_p k_d q_d}\;=\;\frac{1}{\sigma_0}\;Tr(\;T\;\tau_{k_p
q_p}(\frac{1}{2})\;\tau_{k_d q_d}(1)\;T^\dagger\;)
\label{eq:Tkq}
\end{equation}
where $\sigma_0=Tr(TT^\dagger)$ is proportional to the unpolarized cross
 section and $\tau_{k q}(s)$ are the usual spherical spin tensors for spin
 $s$. Choosing the $z$ axis along $\vec{k}$, the invariance under rotations
 along this axis implies that the non-vanishing $T_{k_p q_p k_d q_d}$ must
 have $q_p + q_d = 0 $. This leaves five observables: the deuteron analysing
 power $T_{20}=T_{0020},\;T_{1010},\;T_{111-1},\;T_{112-1}$ and
 $\sigma_{unpol}$ (the unpolarized cross section). However only four are
 independent since they are  bilinear forms of just two amplitudes $T_+$ and
 $T_-$:
\begin{equation}
\sigma_0=2(|T_-|^2+|T_+|^2)
\label{eq:sigtkqa}
\end{equation}

\begin{equation}
\sigma_0\;T_{20}=2\sqrt{2}(\frac{1}{2}|T_-|^2-|T_+|^2)
\label{eq:sigtkqb}
\end{equation}

\begin{equation}
\sigma_0\;T_{1010}=-\sqrt{\frac{3}{2}}|T_-|^2
\label{eq:sigtkqc}
\end{equation}

\begin{equation}
\sigma_0\;T_{111-1}=2\sqrt{3}Re(T_+T_-)
\label{eq:sigtkqd}
\end{equation}

\begin{equation}
\sigma_0\;T_{112-1}=i2\sqrt{3}Im(T_+T_-)
\label{eq:sigtkqe}
\end{equation}
In fact from eqs(\ref{eq:sigtkqa}),(\ref{eq:sigtkqb}) and (\ref{eq:sigtkqc})
 we get the relation:
\begin{equation}
T_{20}=-(\sqrt{2}+\sqrt{3}T_{1010})
\label{eq:t20}
\end{equation}
which can be useful to check that the two protons are in a singlet state,
 and the colinearity of the experiment. By measuring $\sigma_{unpol}$,
 $T_{20}$, $T_{111-1}$, and $T_{112-1}$ we completely determine the matrix
 elements $T_+$ and $T_-$ except for an overall phase factor. This is a
 very rare situation which results from the peculiar spin structure of the
 reaction
 $1\oplus\frac{1}{2}\Rightarrow\frac{1}{2}$ and the adopted geometry.\\

We represent the initial state in momentum space by
\begin{equation}
|\vec{k};\sigma_d,\sigma_1>=\sum_{m=-1}^{1}\Phi_{\sigma_d}^m
(\vec{p})|\vec{k};1m,\sigma_1>
\label{eq:initial}
\end{equation}
where
\begin{equation}
\Phi_{\sigma_d}^m(\vec{p})=\sum_{L=0,2}u_L(p)Y_L^\lambda(\hat{p})
 (L\lambda 1m|1\sigma_d)
\label{eq:phi}
\end{equation}
and $u_0$ and $u_2$ represent the deuteron $S$ and $D$ radial wave
 functions. A colinear experiment is sensitive to the deuteron amplitudes
 where the nucleons, in the deuteron rest frame, have either the same
 helicity
\begin{equation}
\phi^d_+(p\vec{e}_z)=\Phi_0^0(p\vec{e}_z)=\frac{1}{\sqrt{4\pi}}
[u_0(p)-\sqrt{2}u_2(p)]
\label{eq:phi0}
\end{equation}
or opposite helicities
\begin{equation}
\phi^d_-(p\vec{e}_z)=\Phi_{-1}^{-1}(p\vec{e}_z)=\frac{1}
{\sqrt{4\pi}}[u_0(p)+\frac{1}{\sqrt{2}}u_2(p)]
\label{eq:phi-}
\end{equation}

Following the notation of ref.\cite{pe90}, we now introduce the quantities
 ${\cal A}^{NN}_{L_{f}m\sigma_1}(\vec{k}_f,\vec{k}_i)$, which are linear
 combinations of the half-off-shell NN T-matrix elements

\begin{equation}
{\cal A}^{NN}_{L_{f}m\sigma_1}(\vec{k}_f,\vec{k}_i)=\sum_{L_i}
<L_f,\vec{k}_f|T^{NN}|L_i,\vec{k}_i><L_i|1m,\sigma_1>
\label{eq:a}
\end{equation}
In the above equation $\vec{k}_i(\vec{k}_f)$ is the initial(final) momentum
 of the interacting pair and $|L_{i,f}>$ represent the eight orthogonal
 3-particle spin states $|s_{pp}SM>$, defined in ref.\cite{pe90}.
 By detecting the two protons in coincidence we only have to consider
 $L_f=7,8$, where $|7>=|0\frac{1}{2}\frac{1}{2}>$ and
 $|8>=|0\frac{1}{2}-\frac{1}{2}>$.\\

The impulse approximation has been successfully applied to the analysis of
 $d(p,2p)n$ reactions at intermediate energies and for small  recoil
 momentum of the neutron ($q_n < 200MeV/c$)
 \cite{bu85,pu88,mo89,ko91,ca91,pe90}.
 Within this approximation the $T$ matrix in a colinear experiment is given
  by (we neglect the binding energy of the deuteron)
\begin{equation}
T^{IA}_\pm(0^\circ)=\sqrt{2}[{\cal A}^{pp}_\pm(0,-\frac{\vec{k}}{2})\phi_
\pm^d(\frac{\vec{k}}{2})+{\cal A}_\pm^{np}(-\frac{3}{4}\vec{k},
-\frac{5}{4}\vec{k})\phi_\pm^d(\vec{k})]
\label{eq:T+-0}
\end{equation}
for $\vec{k}'=\vec{k}$ and
\begin{equation}
T^{IA}_\pm(\pi)=\sqrt{2}[{\cal A}_\pm^{pp}(0,-\frac{3}{2}\vec{k})
\phi_\pm^d(\frac{3}{2}\vec{k})+{\cal A}_\pm^{np}(\frac{3}{4}\vec{k},
-\frac{3}{4}\vec{k})\phi_\pm^d(0)]
\label{eq:T+-pi}
\end{equation}
for $\vec{k}'=-\vec{k}$. In eqs(\ref{eq:T+-0},\ref{eq:T+-pi})
${\cal A}_+^{NN}(\vec{k}_f,\vec{k}_i)={\cal A}^{NN}_{70\frac{1}{2}}
(\vec{k}_f,\vec{k}_i)$ and  ${\cal A}_-^{NN}(\vec{k}_f,\vec{k}_i)=
{\cal A}^{NN}_{8-1-\frac{1}{2}}(\vec{k}_f,\vec{k}_i)$.\\

Calculations of the helicity amplitudes using deuteron wave functions
 generated by various realistic NN interactions show that they vanish for
 high momenta:
\begin{equation}
\phi_-^d(\vec{p})=0
\label{eq:phi-=0}
\end{equation}
\begin{equation}
\phi_+^d(\vec{p})=0
\label{eq:phi+=0}
\end{equation}
The node in $\phi_-^d(\vec{p})$ is strongly correlated to the node in the
 S-state wave function $u_0(p)$ and results from the decrease of the radial
 function before it changes sign. We find that the solution of
 eq(\ref{eq:phi-=0}) occurs for $1.4fm^{-1}<p<1.7fm^{-1}$ and is quite
 sensitive to the NN interaction used \cite{re68}-\cite{bo89} to generate
 the deuteron wave function. This is shown in Fig.1, where
 $\sqrt{8\pi}\phi_-^d(p)$ is plotted in the region of interest.
 $\phi_+^d(\vec{p})$ is also expected to vanish, but at considerably
 higher momentum, near $5fm^{-1}$. However the predicted solutions are
 much less reliable since at these high momenta non-nucleonic degrees
 of freedom are important.\\

The most important term in eq.(\ref{eq:T+-0}) is expected to be the one
 involving the $T^{pp}$ matrix element\cite{pu88,pu89,pe90}. Neglecting the
 latter, $T^{IA}_\pm(0^\circ)$ becomes proportional to $\phi_\pm^d$ and
 therefore changes sign at the solution of
 eqs(\ref{eq:phi-=0},\ref{eq:phi+=0}). We then conclude that the main
 component of the total T-matrix element $T_-$ ($T^{IA}_-$) changes sign
 for a momentum close to $3.fm^{-1}$, leading to the following two facts:
 i) in this region of momenta rescattering terms become competitive;
 ii) at a certain value of momentum $T^{IA}_-$ will be exactly symmetric
 to the remaining terms, and the total matrix element $T_-$ will have a
 node; its location depends strongly on the NN interaction used and as well
 as on the rescattering terms. The vanishing of $T_-$ implies
\begin{equation}
T_{20}(0^\circ)=-\sqrt{2}
\label{eq:T20min}
\end{equation}
and the spin transfer coefficients all vanish. $T_{20}$ reaches its minimum
 possible value because the polarized cross section for aligned deuterons
 along the $z$ axis vanishes. \\

The present results show that in the $\vec{d}(\vec{p},(pp)_0)n$ reaction
 with a colinear geometry there are only two independent T-matrix elements.
 Furthermore these matrix elements can be determined directly from
 measurements of the unpolarized cross section and spin correlation
 observables. This fact presents a considerable challenge to the theory
 of the reaction and should provide new information on the NN T-matrix
 elements and the deuteron wave function at high momentum. In particular
 $T_{20}$ is predicted to reach a minimum value for an incoming kinetic
 energy close to 410 MeV in the laboratory frame, i.e. a recoil neutron
 momentum of around $270 MeV/c$. The minimum in $T_{20}$ corresponds
 to a node in $T_-$, which is related to a node in the deuteron
 helicity amplitude $\phi_-^d(\vec{p})$. Analogously, in the
 $\vec{d}(e,e'p)n$ reaction \cite{br92,fr83}, $T_{20}(0^\circ)$
 is also predicted to reach the value $-\sqrt{2}$ for
 $\phi_-^d(\vec{p})=0$, neglecting the e-n scattering amplitude,
 final state interactions and other lower order effects.\\

Useful discussions with W.Haeberli and K.Pitts are gratefully
 acknowledged.

\newpage
\begin{center}
{\bf \large Figure caption}
\end{center}
Fig.1 - The $\sqrt{8\pi}\phi_-^d(p)$ amplitude calculated with different
 NN interactions.


\newpage
\begin{thebibliography}{aaa}
\bibitem{ab83} V.G. Ableer et al., Nucl. Phys. A393 (1983) 491
\bibitem{an83} L. Anderson et al., Phys. Rev. C28 (1983) 1224
\bibitem{bu85} D.V. Bugg and C. Wilkin, Phys. Lett. 152 (1985) 37
\bibitem{pu88} V. Punjabi et al, Phys. Rev. C38 (1988) 2728
\bibitem{pu89} V. Punjabi et al, Phys. Rev. C39 (1989) 608
\bibitem{mo89} T. Motobayashi et al., Phys. Lett. B233 (1989) 69
\bibitem{ko91} S. Kox et al., Phys. Lett. B266 (1991) 264
\bibitem{ca91} J. Carbonell, M.B. Barbaro and C. Wilkin, Nucl. Phys.
A529 (1991) 653
\bibitem{zb92} I. Zborovsky, Z.Phys. A343 (1992) 34
\bibitem{ch92} E. Cheung et al, Phys. Lett. 284B (1992) 210
\bibitem{pe90} C.F. Perdrisat and V.Punjabi, Phys. Rev. C42 (1990) 1899
\bibitem{re68} R.V. Reid, Ann. Phys. 50 (1968) 411
\bibitem{ur81} I.E. Lagaris and V.R. Pandharipande, Nucl. Phys.
A359 (1981) 331
\bibitem{ar84} R.B. Wiringa et al., Phys.Rev. C29 (1984) 1207
\bibitem{pa80} M. Lacombe at al., Phys. Rev. C21 (1980) 861
\bibitem{bo89} R. Machleidt et al., Adv. Nucl. Phys. 19 (1989) 189:
Bonn-energy dependent and Bonn1-energy independent
\bibitem{br92} J.F.J. van den Brand, Nucl. Phys. A546 (1992) 299
\bibitem{fr83} Frankfurt et al., Nucl. Phys. A405 (1983) 557


\end{thebibliography}
\end{document}